\documentclass[conference]{IEEEtran}
\IEEEoverridecommandlockouts
\usepackage{cite}
\usepackage{amsmath,amssymb,amsfonts}
\usepackage{algorithmic, subcaption}
\usepackage{graphicx}
\usepackage{textcomp}
\usepackage{xcolor}
\def\BibTeX{{\rm B\kern-.05em{\sc i\kern-.025em b}\kern-.08em
    T\kern-.1667em\lower.7ex\hbox{E}\kern-.125emX}}
\begin{document}

\title{Spiking Neural Networks for Resource Allocation in UAV-Enabled Wireless Networks\\
\thanks{The work of V. Kouvakis and S. E. Trevlakis was funded by the European Union's HORIZON-JU-SNS-2022 Research and Innovation Programme under Grant 101096456. The work of A.-A. A. Boulogeorgos was supported by the project MINOAS. The research project MINOAS is implemented in the framework of H.F.R.I call ``Basic research Financing (Horizontal support of all Sciences)'' under the National Recovery and Resilience Plan “Greece 2.0” funded by the European Union – NextGenerationEU (H.F.R.I. Project Number: 15857).}}

\author{\IEEEauthorblockN{Vasileios Kouvakis\IEEEauthorrefmark{1}\IEEEauthorrefmark{2}, Stylianos E. Trevlakis\IEEEauthorrefmark{1}\IEEEauthorrefmark{4}, Ioannis Arapakis\IEEEauthorrefmark{3}, and Alexandros-Apostolos A. Boulogeorgos\IEEEauthorrefmark{2}
\vspace{0.35mm}
\IEEEauthorblockA{\IEEEauthorrefmark{1}Department of Research and Development, InnoCube P.C., 17th Noemvriou 79, Thessaloniki 55534, Greece.\\
\IEEEauthorrefmark{2}Department of Electrical and Computer Engineering, University of Western Macedonia, Kozani 50100, Greece.\\
\IEEEauthorrefmark{3}Telefonica Scientific Research, Barcelona, Spain.\\ 
\IEEEauthorrefmark{4}Department of Informatics \& Telecommunications, University of Thessaly, Lamia 35100, Greece.\\
\vspace{0.35mm}
\IEEEauthorblockA{Emails: \{kouvakis, trevlakis\}@innocube.org, ioannis.arapakis@telefonica.com, al.boulogeorgos@ieee.org
}}}}

\maketitle

\begin{abstract}
This work presents a new spiking neural network (SNN)-based approach for user equipment-base station (UE-BS) association in non-terrestrial networks (NTNs). With the introduction of UAV's in wireless networks, the system architecture becomes heterogeneous, resulting in the need for dynamic and  efficient management to avoid congestion and sustain overall performance. The presented framework compares two SNN-based optimization strategies. Specifically, a top-down centralized approach with complete network visibility and a bottom-up distributed approach for individual network nodes. The SNN is based on leak integrate-and-fire neurons with temporal components, which can perform fast and efficient event-driven inference. Realistic ray-tracing simulations are conducted, which showcase that the bottom-up model attains over 90\% accuracy, while the top-down model maintains 80-100\% accuracy. Both approaches reveal a trade-off between individually optimal solutions and UE-BS association feasibility, thus revealing the effectiveness of both approaches depending on deployment scenarios.
\end{abstract}


\section{Introduction}
For the advancement of state-of-the-art networks, innovative strategies in mobility management are required to cope with the data surge as well as the variety of available services~\cite{Koutsonas2023}. Since next generation networks aim at improving connectivity through non-terrestrial networks (NTNs) and heterogeneous networks (HetNets), proper association between user equipment devices (UEs) and base stations (BSs) becomes critical for ensuring high system performance, ensuring quality of service guarantees, and avoiding congestion and interference. Utilizing artificial intelligence (AI) for solving the UE-BS association problem improves the achievable throughput, increases accuracy, and increases the network adaptability, all of which are important in ultra-dense networks~\cite{Ponnusamy2022}. 

The importance of adaptive association of users and BSs comes from maximizing network throughput, improving energy consumption, and lowering carbon footprints~\cite{Trevlakis2024}. AI algorithms can modify UE-BS associations in real-time to improve load balancing and efficiency~\cite{Lu2022,Agarwal2023}. Adaptive user association enables renewable energy powered BSs to operate with the maximum number of connected users, which can provide important reduction of the network's carbon emissions~\cite{Liu2014}. Approaches that integrate user association with resource allocation can improve energy efficiency further by ensuring that BSs operate at their optimal power margins. In dense networks, especially at higher frequencies, more advanced methods of enabling users to connect to several BSs simultaneously can reduce interference as well as improve network performance showcasing user experience improvements~\cite{Lin2019,Ebrahim2017}.

Spiking Neural Networks (SNNs) and Reconfigurable Intelligent Surfaces (RIS) have advanced the domains of AI and wireless communication, respectively. SNNs imitate biological neurons by utilizing energy-efficient event-driven processing. This allows for complex computations to be performed using rank ordering through parameters reconfiguration~\cite{Kouvakis2025}. RIS technology, on the other hand, uses reconfigurable elements and signal adaptive techniques to manipulate wireless signals for extending network coverage or providing improved connectivity in areas with mediocre signal quality~\cite{Boulogeorgos}. The combination of the two technologies has high potential for communication systems. Implementing SNNs can increase the dynamic character of RISs, allowing the creation of intelligent systems that adjust to users in real-time.

Despite the growing interest in AI-driven solutions for NTN optimization, existing literature lacks comprehensive approaches that leverage the temporal dynamics and energy efficiency of SNNs for UE-BS association problems in UAV-enabled heterogeneous networks. This paper addresses this gap by presenting the first SNN-based framework specifically designed for dynamic UE-BS association in NTN environments. The main technical contributions include: 
\begin{itemize}
    \item The development of a novel SNN architecture employing leaky integrate-and-fire (LIF) neurons with temporal processing capabilities for real-time UE-BS association decisions.
    \item The formulation and comparison of two distinct optimization strategies. Specifically, a centralized top-down approach with complete network visibility and a distributed bottom-up approach for individual network nodes.
    \item The integration of realistic operational constraints into the SNN training process through constraint-aware loss function design.
    \item A comprehensive performance evaluation using realistic ray-tracing simulations in urban environments, demonstrating that the bottom-up model achieves over $90$\% accuracy while the top-down model maintains $80-100$\% accuracy with superior constraint satisfaction.
\end{itemize}

\section{Methodology}
This section presents the SNN-based intelligent UE-BS association approach for dynamically evolving UAV-enabled NTN systems. Specifically, it covers problem formulation, dataset creation, data preprocessing, SNN design, and training methodology. 

\subsection{System Model}
We assume that an urban geographic area is covered by two BSs and one UAV-mounted RIS that has been deployed ad-hoc for improving the coverage of the network. Both BSs are equipped with a directional antennas based on the 3GPP TR38901 technical specification~\cite{3GPP_TR38901_2017}, while the UAV-mounted RIS is considered as an ideal phase gradient reflector. Moreover, the area under investigation is populated with multiple users that are assumed to be moving randomly in the environment, with each user is equipped with a UE that houses a unidirectional receiving antenna. 

\begin{figure}
    \centering
    \includegraphics[width=1\linewidth]{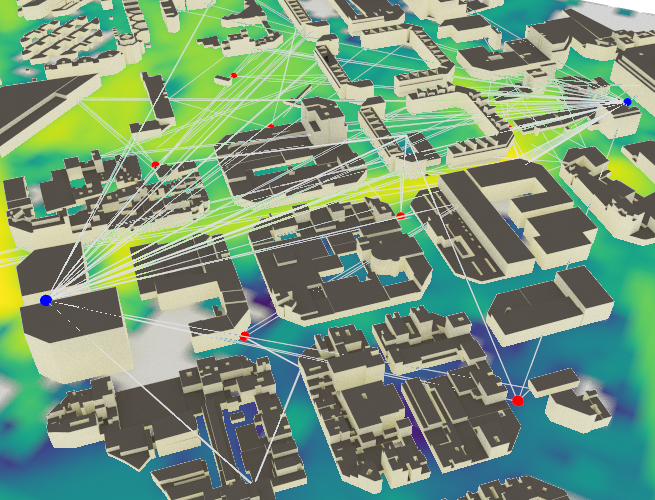}
    \caption{3D representation of the considered system model.}
    \label{Fig:3D_scene}
\end{figure}

The training dataset was generated through realistic ray-tracing wireless communication scenario simulations to ensure that the resulting data accurately reflects real-world channel conditions and network dynamics. The Sionna library was utilized for performing the ray-tracing simulations~\cite{sionna}. The simulation framework models various aspects of wireless propagation, including path loss, reflections, diffraction, shadowing effects, and multipath fading~\cite{Crysovergis2025}. A three dimensional representation of the simulated setup is provided~in~Fig.~\ref{Fig:3D_scene}. The UEs move with a random walk algorithm. A real-world map of Barcelona was used for the simulation. It was imported in Blender~\cite{blender} through the OpenStreatMap API~\cite{openstreetmap}. In addition, real-world positions for the BSs were selected from the CellMapper~\cite{cellmapper}.

\begin{figure}
    \centering
    \includegraphics[width=1\linewidth]{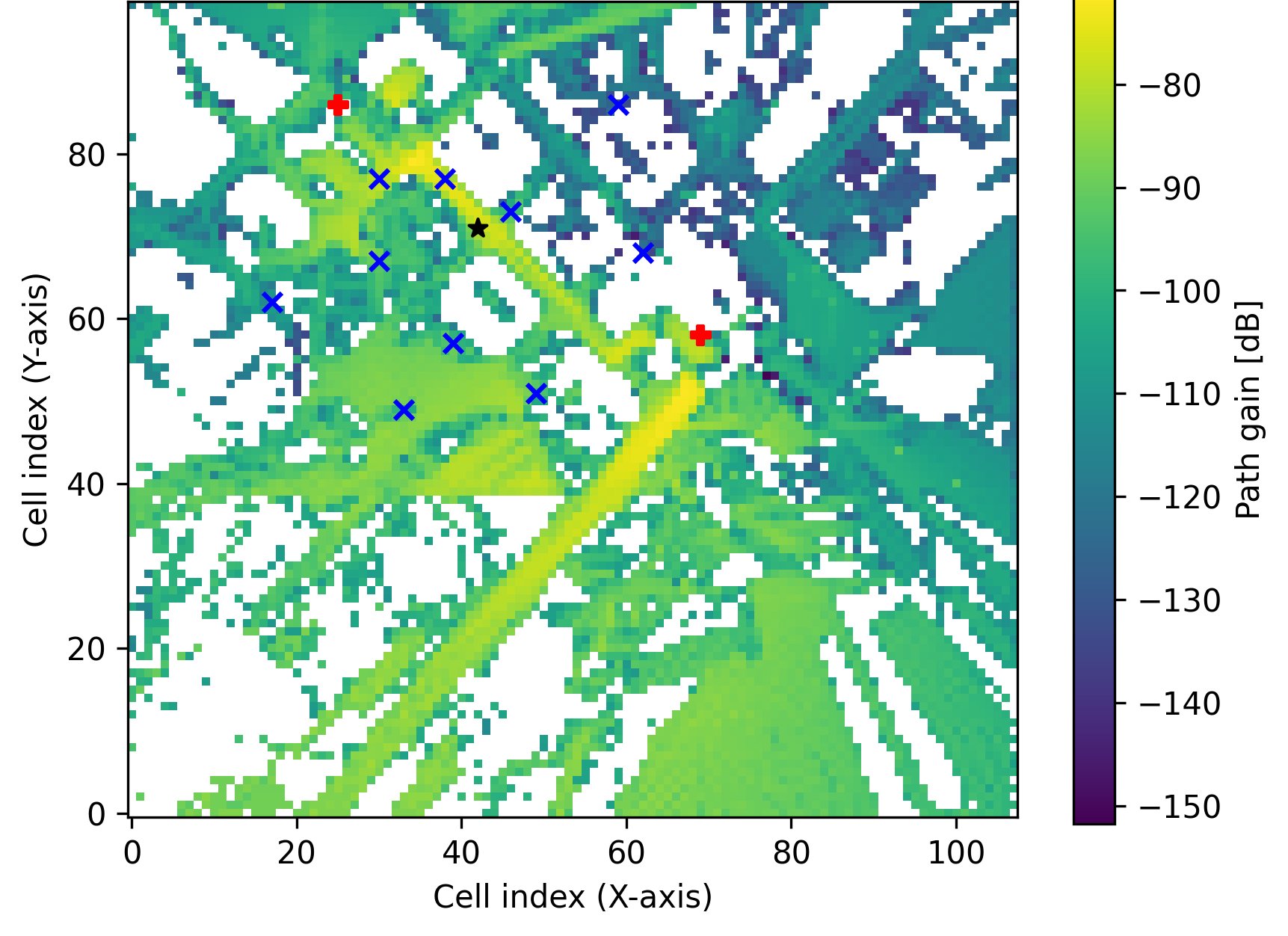}
    \caption{Coverage map.}
    \label{Fig:coverage_map}
\end{figure}

\begin{figure}
    \centering
    \includegraphics[width=1\linewidth]{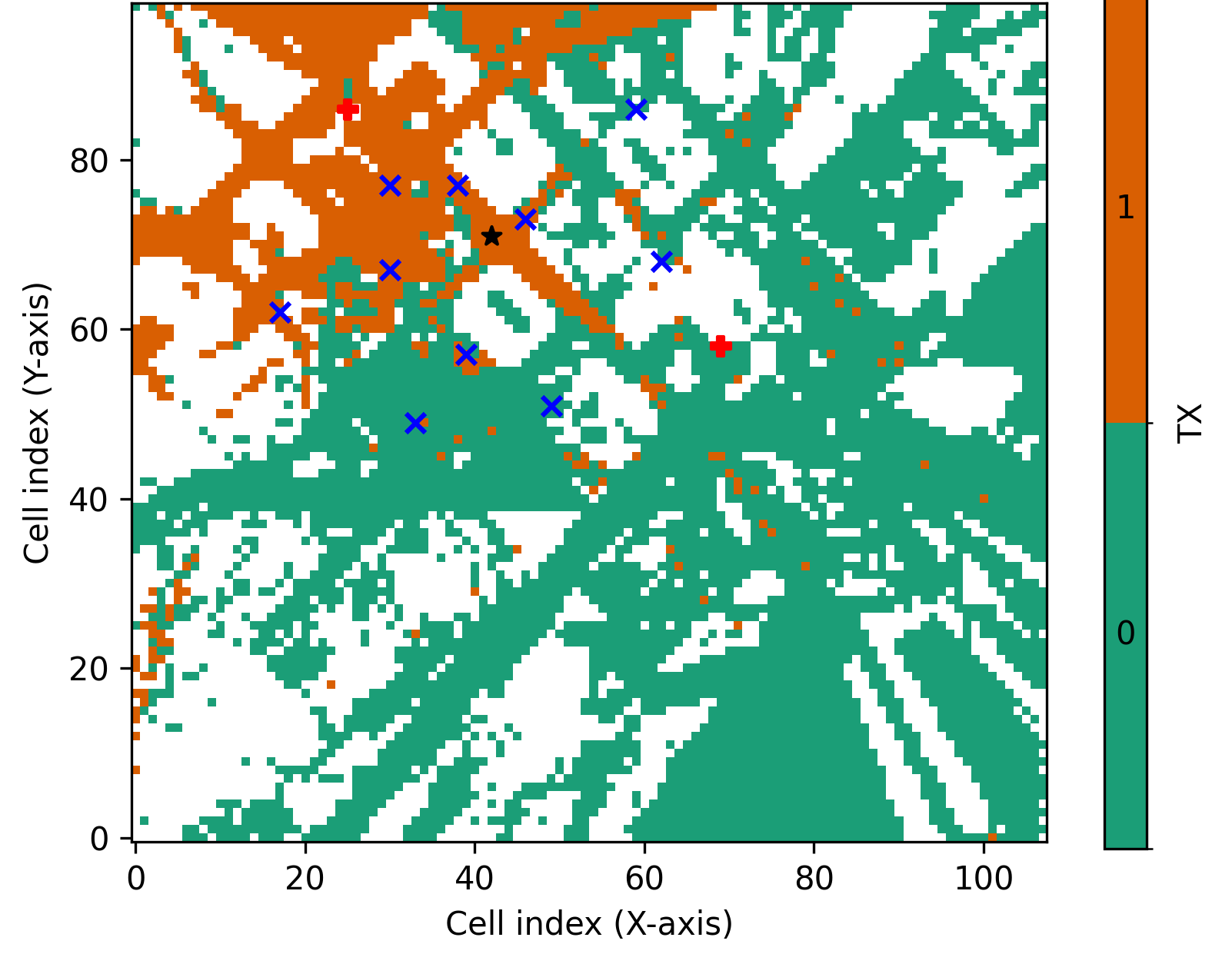}
    \caption{UE-BS association.}
    \label{Fig:association}
\end{figure}

The simulations were carried out in distinct time steps. Each dataset instance contains the complete data rate matrix representing all possible UE-BS associations within the network scenario. The coverage map and the UE-BS associations of a single time step are visualized~in~Fig.~\ref{Fig:coverage_map} and~Fig.~\ref{Fig:association}, respectively. This comprehensive representation enables the learning algorithm to capture the full complexity of the association problem, including interdependencies between different RX associations and their collective impact on network performance. 
Ground truth optimal associations were computed using brute force optimization, exhaustively evaluating all possible association combinations to identify the configuration that achieves maximum total network data rate while satisfying the operational constraints. This approach ensures that the training targets represent true optimal solutions, providing a reliable benchmark for model performance evaluation. The input features undergo preprocessing to normalize the data rate values across different network scenarios. Feature scaling ensures numerical stability during training and prevents any single UE-BS pair from dominating the learning process due to scale differences. The standardization process maintains the relative relationships between data rates while ensuring that the neural network can effectively process the input patterns.

\subsection{Problem Formulation}
The UE-BS association problem in wireless communication networks is a combinatorial optimization challenge that aims to maximize the total data rate network throughput while satisfying realistic operational constraints. Given a set of receivers (RXs) and transmitters (TXs), each UE-BS pair exhibits a specific data rate based on channel conditions, interference patterns, and propagation characteristics. The objective is to determine the optimal association of each RX to a TX or RIS such that the aggregate network data rate is maximized. Let $R = {r_1, r_2, ..., r_N}$ represent the set of $N$ RXs and $T = {t_1, t_2, ..., t_M}$ represent the set of $M$ TXs in the network. For each UE-BS pair $(r_i, t_j)$, we define $d_{i,j}$ as the achievable data rate between RX $i$ and TX $j$. The association problem seeks to find a mapping function $f: R \rightarrow T$ that maximizes the total network throughput denoted by
\begin{align}
    &\max \quad\sum_{i=1}^{N} d_{i,f(r_i)} \\
    &\quad \textrm{s.t.} \ |R \rightarrow t_i| \leq L
\end{align}
where $|R \rightarrow t_i|$, which denotes the numbed of RXs assigned to a specific TX, $t_i$, cannot exceed the threshold $L$. This constraint reflects practical limitations in wireless systems, including hardware capabilities, power limitations, and interference management requirements. The constraint ensures that UE-BS associations remain feasible under real-world deployment scenarios. The combinatorial nature of this optimization problem leads to exponential complexity as the number of RXs and TXs increases. Traditional exhaustive search methods become computationally prohibitive for large-scale networks, motivating the development of efficient AI-based approaches that can provide near-optimal solutions with significantly reduced computational overhead. 

\subsection{Spiking Neural Network} 
The choice of SNNs for the UE-BS association problem is motivated by several key advantages that align with the requirements of wireless communication systems. SNNs operate based on temporal dynamics and event-driven processing, incorporating the concept of time as a fundamental computational element. This temporal processing capability is particularly valuable for wireless systems where timing considerations and dynamic channel conditions play crucial roles. The event-driven nature of SNNs enables selective neuron activation only when input signals are significant enough to trigger spikes, leading to improved power consumption and computational resource efficiency. This characteristic is especially beneficial for deployment scenarios where energy efficiency is critical, such as battery-powered base stations or edge computing environments. The sparse activation patterns in SNNs naturally align with the sporadic nature of network optimization requirements, where associations may only need updates when channel conditions change significantly. Furthermore, SNNs can capture complex temporal patterns and dependencies in the data that traditional feedforward networks might miss. In the context of UE-BS association, these temporal dynamics can help model the evolving nature of channel conditions and the interdependencies between different association decisions over time. 
\begin{figure}
    \centering
    \includegraphics[width=1\linewidth]{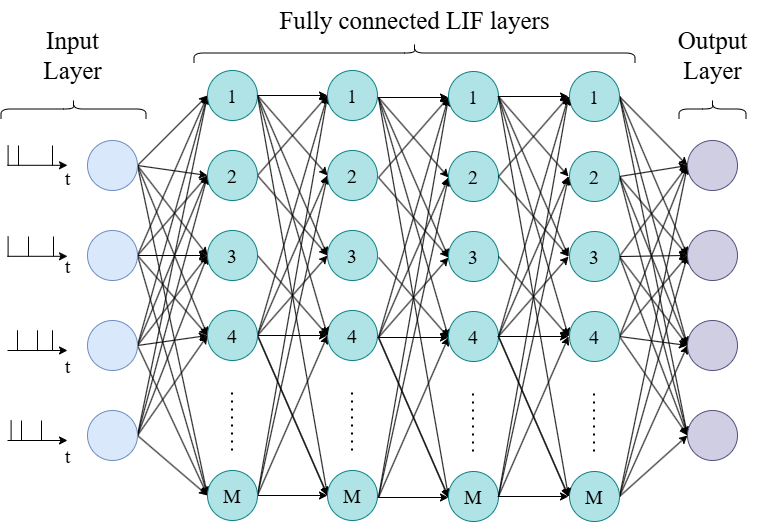}
    \caption{SNN architecture.}
    \label{Fig:architecture}
\end{figure}

\subsubsection{Architecture Design} 
The adopted SNN architecture, as illustrated~in~Fig.~\ref{Fig:architecture}, employs leaky integrate-and-fire (LIF) neurons as the fundamental computational units. LIF neurons maintain membrane potentials that integrate incoming currents over time and generate spikes when the potential exceeds a threshold. The leaky characteristic allows the membrane potential to gradually decay, providing a natural forgetting mechanism that prevents indefinite accumulation of past inputs. The network consists of multiple fully connected layers with LIF neurons, creating a deep architecture capable of learning complex non-linear mappings between input data rates and optimal associations. Each layer incorporates dropout regularization to prevent overfitting and improve generalization performance. The dropout mechanism randomly deactivates a fraction of neurons during training, encouraging the network to develop robust internal representations that do not rely on specific neuron activations. The input layer receives the flattened data rate matrix representing all possible UE-BS associations. The hidden layers progressively transform these features through multiple levels of abstraction, enabling the network to learn both local patterns within individual UE-BS pairs and global dependencies across the entire network. The output layer produces association probabilities for each RX, with the dimensionality matching the number of possible UE-BS combinations. The temporal aspect of the SNN is implemented through iterative processing over multiple time steps. During each time step, the same input is presented to the network, allowing the LIF neurons to accumulate and process information temporally. The membrane potentials of the output neurons are recorded across all time steps, with the final membrane potentials used for association prediction.

\subsubsection{Loss Function Design} 
The training objective combines multiple components to address both the optimization goal and the operational constraints. The primary component uses cross-entropy loss to encourage correct association predictions by comparing the predicted probability distributions with the ground truth optimal associations. This component ensures that the network learns to identify the TX choices that maximize data rates for each RX. A constraint penalty term is incorporated to discourage solutions that violate the realistic operational limits on the number of RXs per TX. The penalty weight is calibrated to provide sufficient constraint enforcement without overwhelming the primary optimization signal, thus guiding the network toward feasible solutions while maintaining focus on the primary optimization objective. The combined loss function balances these competing objectives through careful weighting, ensuring that the network learns to produce high-quality associations that are both near-optimal in terms of data rate and feasible under realistic constraints.

\subsection{Integration Approaches} 
\begin{figure}
    \centering
    \includegraphics[width=1\linewidth]{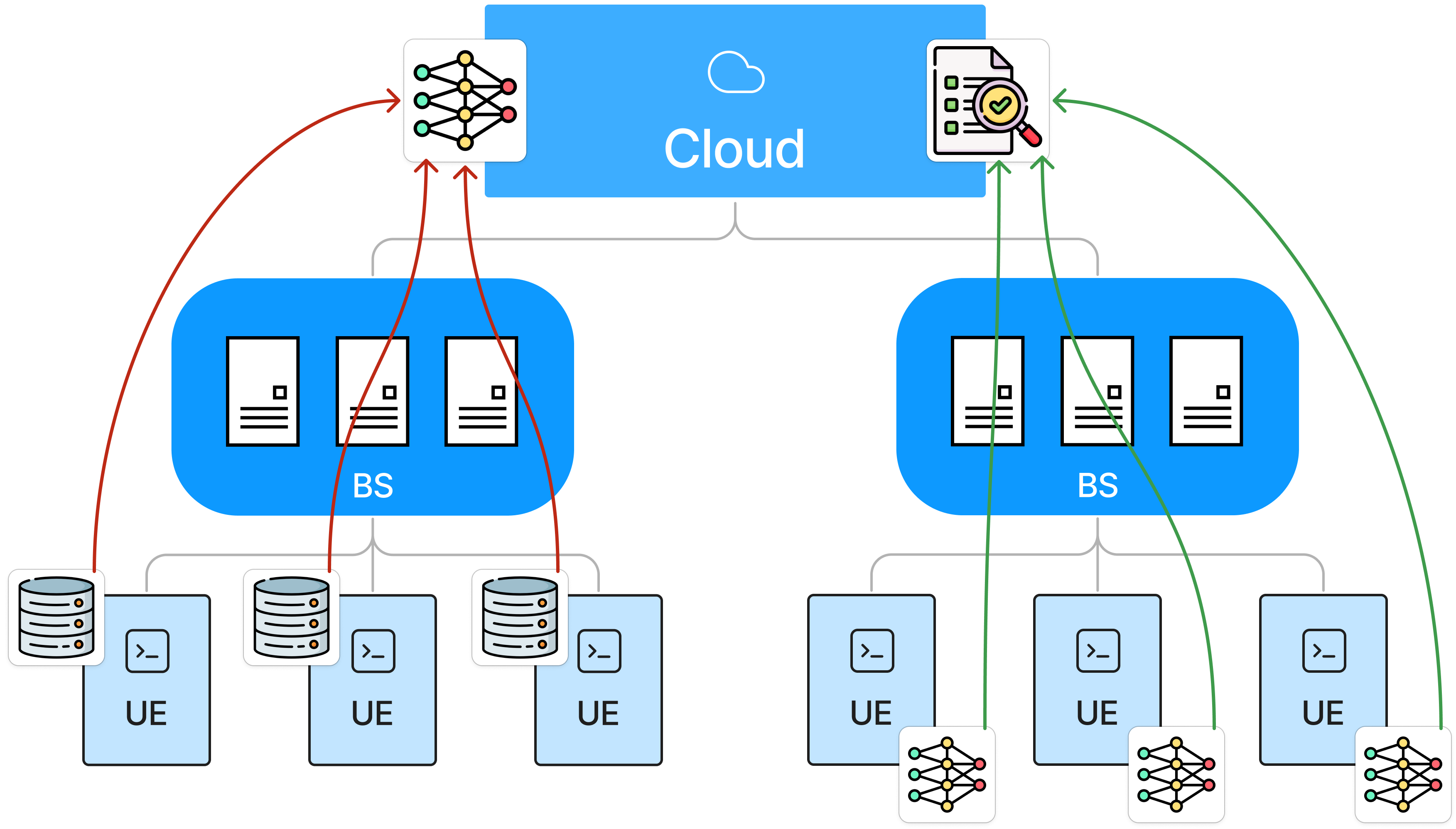}
    \caption{Top-down vs bottom-up approaches.}
    \label{Fig:approaches}
\end{figure}

As~illustrated~in~Fig.~\ref{Fig:approaches}, two distinct approaches have been identified as promising for solving the association problem, namely top-down and bottom-up.

\subsubsection{Top-Down Approach} 
The top-down model represents a centralized approach to the UE-BS association problem. The top-down training approach presents the complete network state as input during each training iteration. This comprehensive view allows the model to understand how individual association decisions affect other RXs and TXs in the entire network. The model learns to identify scenarios where certain association choices create cascading effects on network performance. Its ability to access complete network information enables it to learn and enforce realistic constraints effectively. The training process for the top-down model involves gathering information from all network nodes at each simulated time step and transmitting this information to the centralized location where the SNN model is located.

\subsubsection{Bottom-Up Approach}  
The bottom-up model represents a distributed approach suitable for deployment at individual UEs where only limited network information is available. Unlike the top-down model, it disregards the connections of other UEs. This approach eliminates the model's ability to consider cross-RX dependencies and global constraint satisfaction. The bottom-up training methodology uses the same dataset as the top-down model but restructures the training process to focus on individual RX decisions. Both models utilize similar SNN architectures and neuron types, but operate on different input dimensionalities. 

\subsection{Training Process and Optimization} 
The training process employs adaptive learning rate optimization using the Adam algorithm. The learning rate schedule incorporates plateau-based reduction, automatically decreasing the learning rate when validation performance reaches a certain point. This adaptive approach ensures stable convergence while preventing premature termination of the learning process. The dataset is partitioned into training and validation splits using an $80-20$ ratio, ensuring adequate data for both model training and unbiased performance evaluation. The training process monitors both training and validation metrics to detect overfitting and guide hyperparameter adjustments. Batch processing is employed to improve training efficiency and gradient estimation stability. The batch size is selected to balance memory requirements with training stability, enabling effective utilization of available computational resources while maintaining consistent gradient quality across training iterations. The training continues for a predetermined number of epochs, with early stopping mechanisms based on validation performance to prevent overfitting. Model checkpointing preserves the best-performing model configurations throughout the training process, ensuring that the final model represents the optimal balance between training performance and generalization capability. 

\begin{figure}
    \centering
        \begin{subfigure}[b]{1\linewidth}
        \centering
        \includegraphics[width=1\linewidth]{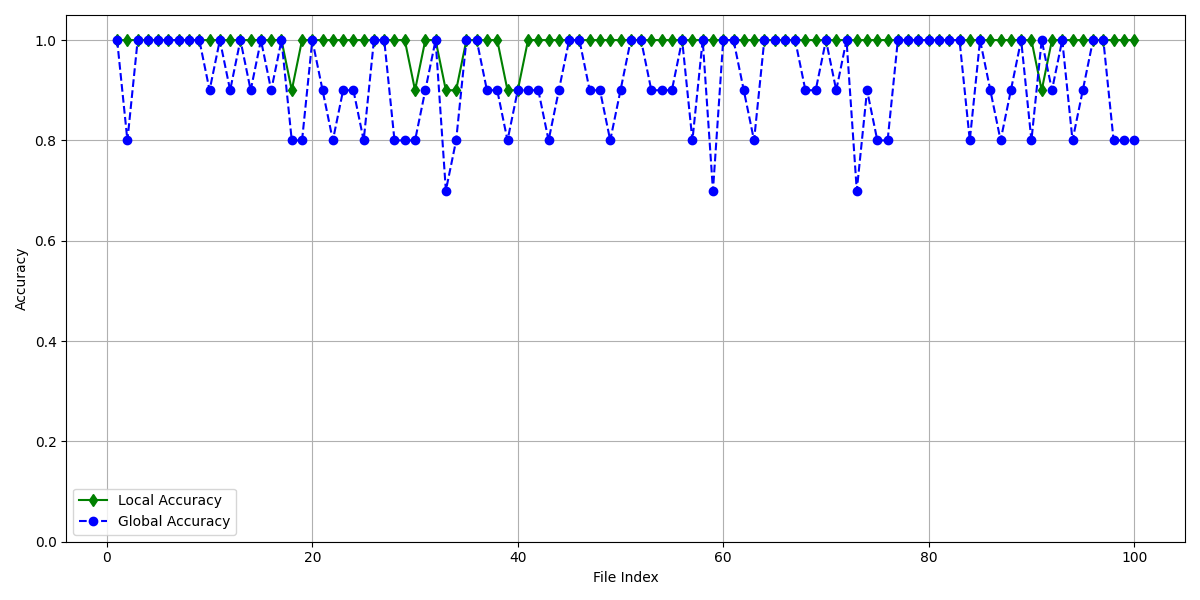}
        \caption{Comparison of the achievable accuracy of the two approaches.}
        \label{Fig:accuracy_merged}
    \end{subfigure}
    \hfill
    \begin{subfigure}[b]{1\linewidth}
        \centering
        \includegraphics[width=1\linewidth]{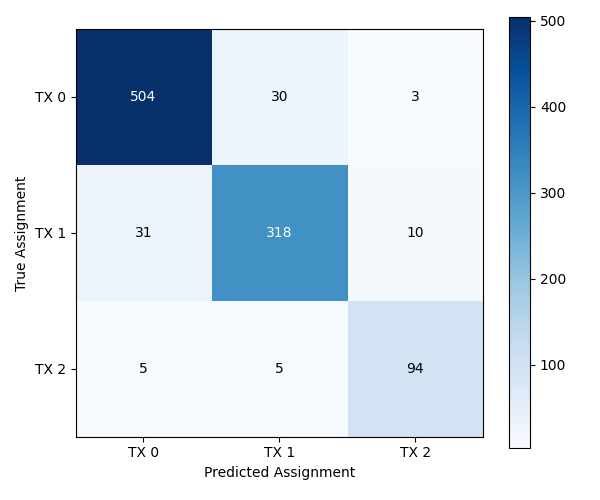}
        \caption{Confusion matrix of the top-down approach.}
        \label{Fig:confusion_global}
    \end{subfigure}
    \hfill
    \begin{subfigure}[b]{1\linewidth}
        \centering
        \includegraphics[width=1\linewidth]{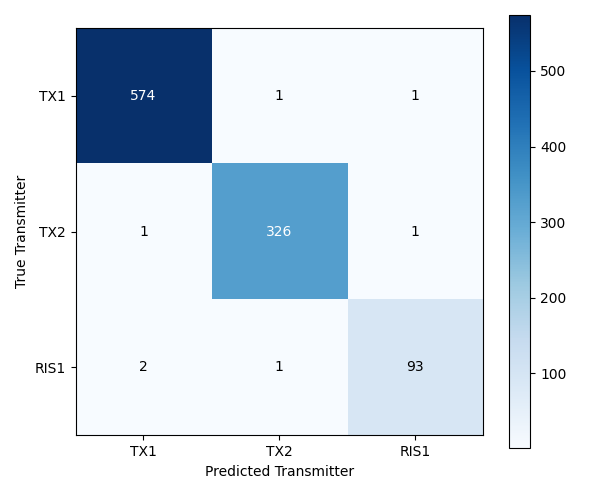}
        \caption{Confusion matrix of the bottom-up approach.}
        \label{Fig:confusion_local}
    \end{subfigure}
    \caption{Comparison of inference performance of the two approaches.}
    \label{fig:inference}
\end{figure}

\section{Results}
Cloud-based high-performance computing resources were used to conduct model development and training. An AWS EC2 g6.4xlarge instance with $16$ vCPUs, $64$ GiB RAM, and $1$ NVIDIA L4 GPU with $22.35$ GiB of memory featured the necessary specifications for large-scale neural network training. The GPU acceleration was particularly beneficial for the iterative temporal loops inherent in SNN computation, significantly reducing training time compared to CPU-only implementations. The evaluation process compares model predictions against the ground truth optimal solutions computed through brute force optimization. This comparison provides an absolute benchmark for assessing solution quality and enables quantitative analysis of the trade-offs between accuracy and efficiency.

\subsection{Inference performance}
Figure~\ref{Fig:accuracy_merged} illustrates the comparative performance of top-down and bottom-up SNN models across $100$ time steps, revealing distinct behavioral patterns that reflect their underlying optimization complexities. The bottom-up model (green) maintains consistently high accuracy above $90$\% throughout the evaluation period, demonstrating superior stability due to its simplified single-RX optimization task that operates without network-wide constraint considerations. In contrast, the top-down model (blue) exhibits significantly higher variability with accuracy fluctuating across an $80-100$\% range, reflecting the inherent difficulty of balancing individual RX optimality with network-wide constraint satisfaction requirements. The top-down model's accuracy fluctuations indicate ongoing constraint-optimization trade-offs. Despite these differences, both models achieve acceptable performance levels above $80$\%, demonstrating the fundamental effectiveness of the SNN architecture for the UE-BS association problem regardless of complexity level.

The confusion matrices of the two models, which are presented~in~Fig.~\ref{Fig:confusion_global}~and~\ref{Fig:confusion_local}, reveal a fundamental trade-off between individual optimization accuracy and network-wide constraint satisfaction. The bottom-up model demonstrates significantly higher individual accuracy with nearly perfect diagonal dominance, achieving $574$, $326$, and $93$ correct predictions for the two TXs and the UAV-mounted RIS, respectively. In contrast, the top-down model exhibits higher confusion between TX choices, with higher misclassification rates, such as approximately $30$ predictions misclassified between TXs. This reflects the inherent complexity of constraint-aware optimization where individually optimal choices may conflict with network-wide requirements. While the bottom-up model's superior individual performance stems from optimizing each RX independently without constraint awareness, the top-down model's confusion patterns indicate realistic trade-offs where misclassifications likely represent scenarios where sub-optimal individual choices enable better overall network performance and constraint satisfaction.

\subsection{Communication performance}
\begin{figure}
    \centering
        \begin{subfigure}[b]{1\linewidth}
        \centering
        \includegraphics[width=1\linewidth]{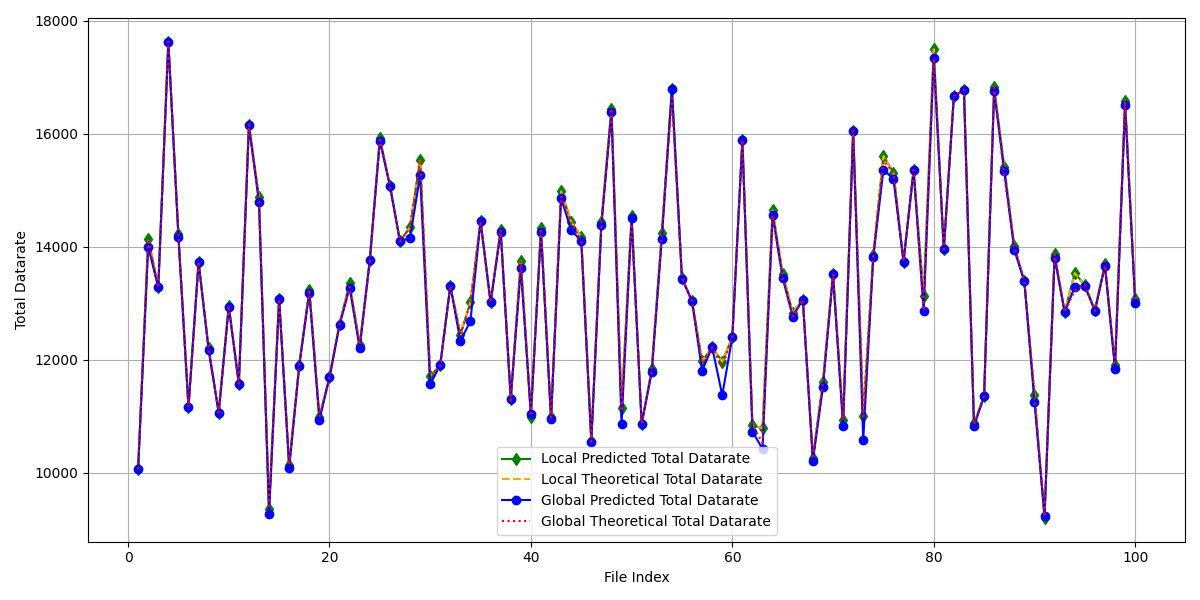}
        \caption{Comparison of the achievable data rate of the two approaches.}
        \label{Fig:datarate_merged}
    \end{subfigure}
    \hfill
    \begin{subfigure}[b]{1\linewidth}
        \centering
        \includegraphics[width=1\linewidth]{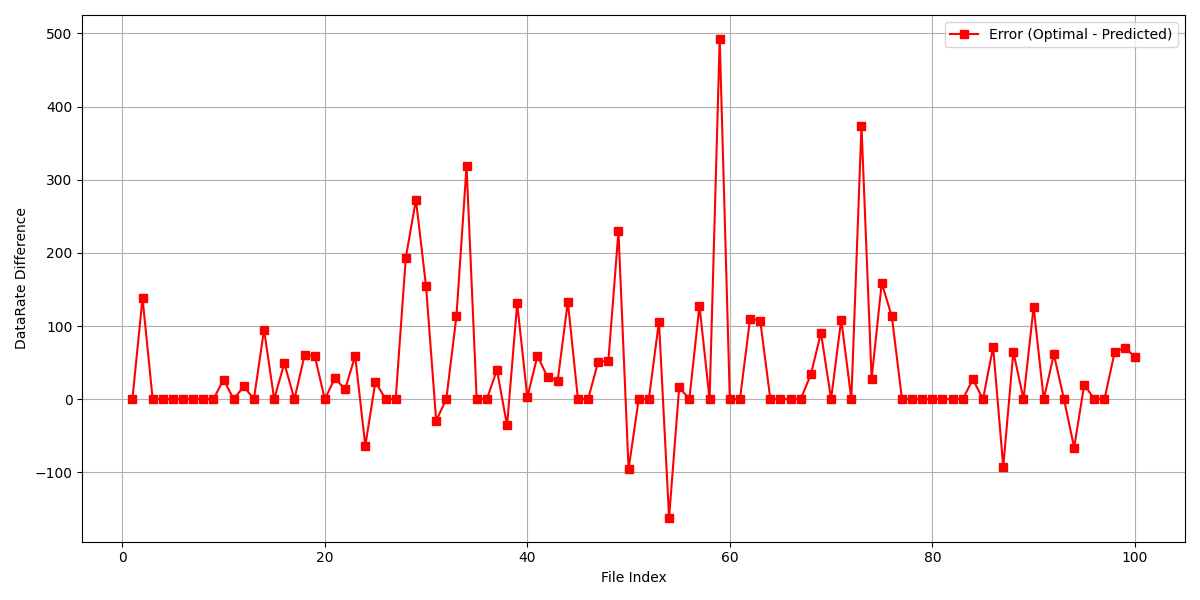}
        \caption{Data rate error of the "top-down" approach.}
        \label{Fig:error_global}
    \end{subfigure}
    \hfill
    \begin{subfigure}[b]{1\linewidth}
        \centering
        \includegraphics[width=1\linewidth]{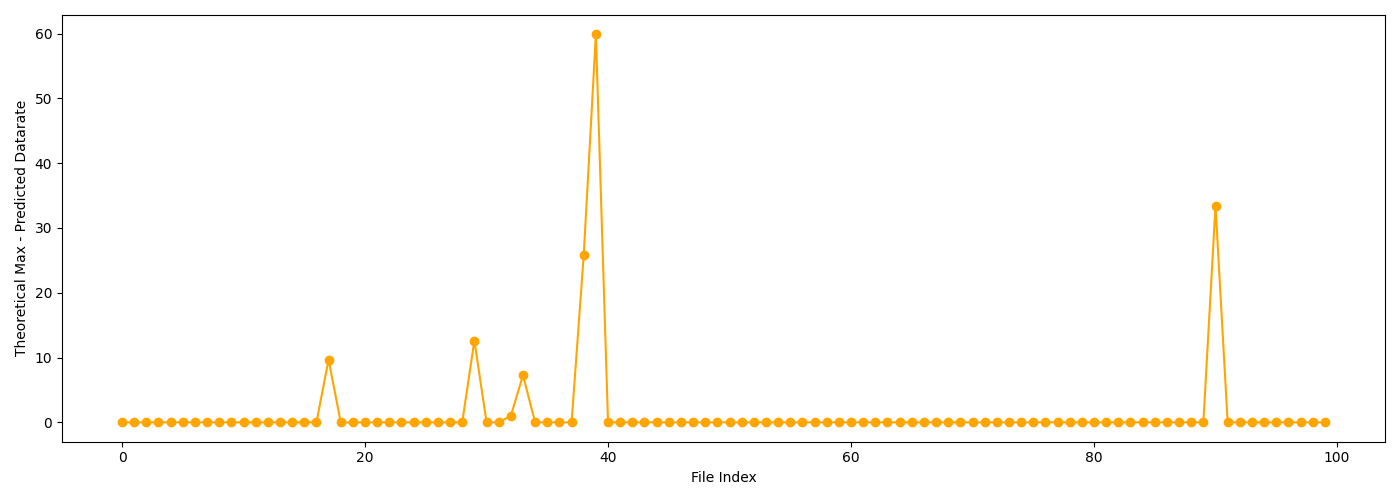}
        \caption{Data rate error of the "bottom-up" approach.}
        \label{Fig:error_local}
    \end{subfigure}
    \caption{Comparison of communication performance of the two approaches.}
    \label{fig:communication}
\end{figure}
Figure~\ref{Fig:datarate_merged} demonstrates the effectiveness of SNNs for the UE-BS association problem, with both bottom-up and top-down models showing strong performance while serving different operational purposes. The bottom-up model (green) achieves near-perfect data rate predictions. The predicted values virtually are identical to theoretical optimal rates across nearly all time steps, exhibiting only minimal deviations in a handful of cases. In contrast, the top-down model (blue) shows consistent but imperfect approximation, maintaining reasonable proximity to optimal rates while operating under more realistic network-wide constraints. Much like the previous figures, the bottom-up model focuses purely on maximizing individual RX data rates, leading to optimal individual decisions. However, the top-down model's constraint-aware approach maintains competitive performance while ensuring realistic network operation. This explains why optimal reference values differ between scenarios.

Figures~\ref{Fig:error_global}~and~\ref{Fig:error_local} present the top-down and bottom-up models' error distribution from optimal performance, respectively. The former showcases several prominent spikes reaching $300-500$ spikes and consistent moderate errors throughout the time steps. In contrast, the latter exhibits with most values at or near zero except for a few notable peaks with deviation of $10-60$. The error distribution plots for both scenarios reveal different optimization strategies where the top-down approach prioritizes system-wide constraints over UE-specific performance maximization.

\section{Conclusions}
This paper details the use of SNNs for calculating the optimal UE-BS association in NTNs under two distinct approaches. Realistic ray-tracing simulations in Barcelona's urban environment showed that the bottom-up approach yielded the highest individual accuracy ($90$\%) in distributed scenarios, while the top-down one performed competitively ($80-100$\%) in centralized deployments. The specialized SNN architecture employing LIF neurons with temporal processing features provides lower energy alternatives to standard neural networks while balancing individual data rate optimization against realistic operational constraints. These findings confirm the feasibility of SNNs for UE-BS association, enabling greater adaptability for network operators while advancing temporal-aware optimization strategies.

\bibliographystyle{IEEEtran}
\bibliography{refs.bib}

\end{document}